\documentclass[manuscript]{aastex631}

\shorttitle{NGC\,1275 flare with MACE}
\shortauthors{Godambe et al.}

\graphicspath{{./}{figures/}}

\begin{document}
\title{Very High Energy Gamma-ray episodic activity of radio galaxy NGC\,1275 in 2022-2023 measured with MACE} 
%\footnote{Released on July, 5th, 2024}}

\correspondingauthor{S. Godambe}
\email{gsagar@barc.gov.in}
\author[0000-0002-0786-7307]{S. Godambe}
\affiliation{Astrophysical Sciences Division, Bhabha Atomic Research Centre \\
Trombay, Mumbai, Maharashtra, India, 400085}

\correspondingauthor{N. Mankuzhiyil}
\email{nijil@barc.gov.in}
\author{N. Mankuzhiyil}
\affiliation{Astrophysical Sciences Division, Bhabha Atomic Research Centre \\
Trombay, Mumbai, Maharashtra, India, 400085}

\correspondingauthor{C. Borwankar}
\email{chinmay@barc.gov.in}
\author{C. Borwankar}
\affiliation{Astrophysical Sciences Division, Bhabha Atomic Research Centre \\
Trombay, Mumbai, Maharashtra, India, 400085}

\author{B. Ghosal}
\affiliation{Astrophysical Sciences Division, Bhabha Atomic Research Centre \\
Trombay, Mumbai, Maharashtra, India, 400085}
\affiliation{Homi Bhabha National Institute, Anushaktinagar, \\
Mumbai, Maharashtra, India, 400094}

\author{A. Tolamatti}
\affiliation{Astrophysical Sciences Division, Bhabha Atomic Research Centre \\
Trombay, Mumbai, Maharashtra, India, 400085}
\affiliation{Homi Bhabha National Institute, Anushaktinagar, \\
Mumbai, Maharashtra, India, 400094}

\author{M. Pal}
\affiliation{Sri Venkateswara College, University of Delhi \\
Benito Juarez Road, Dhaula Kuan, New Delhi, India 110021}
\author{P. Chandra}
\affiliation{Astrophysical Sciences Division, Bhabha Atomic Research Centre \\
Trombay, Mumbai, Maharashtra, India, 400085}

\author{M. Khurana}
\affiliation{Astrophysical Sciences Division, Bhabha Atomic Research Centre \\
Trombay, Mumbai, Maharashtra, India, 400085}
\affiliation{Homi Bhabha National Institute, Anushaktinagar, \\
Mumbai, Maharashtra, India, 400094}

\author{P. Pandey}
\affiliation{Astrophysical Sciences Division, Bhabha Atomic Research Centre \\
Trombay, Mumbai, Maharashtra, India, 400085}

\author{Z. A. Dar}
\affiliation{Astrophysical Sciences Division, Bhabha Atomic Research Centre \\
Trombay, Mumbai, Maharashtra, India, 400085}

\author{S. Godiyal}
\affiliation{Astrophysical Sciences Division, Bhabha Atomic Research Centre \\
Trombay, Mumbai, Maharashtra, India, 400085}

\author{J. Hariharan}
\affiliation{Astrophysical Sciences Division, Bhabha Atomic Research Centre \\
Trombay, Mumbai, Maharashtra, India, 400085}

\author{Keshav Anand}
\affiliation{Astrophysical Sciences Division, Bhabha Atomic Research Centre \\
Trombay, Mumbai, Maharashtra, India, 400085}

\author{S. Norlha}
\affiliation{Astrophysical Sciences Division, Bhabha Atomic Research Centre \\
Trombay, Mumbai, Maharashtra, India, 400085}

\author{D. Sarkar}
\affiliation{Astrophysical Sciences Division, Bhabha Atomic Research Centre \\
Trombay, Mumbai, Maharashtra, India, 400085}
\affiliation{Homi Bhabha National Institute, Anushaktinagar, \\
Mumbai, Maharashtra, India, 400094}

\author{R. Thubstan}
\affiliation{Astrophysical Sciences Division, Bhabha Atomic Research Centre \\
Trombay, Mumbai, Maharashtra, India, 400085}

\author{K. Venugopal}
\affiliation{Astrophysical Sciences Division, Bhabha Atomic Research Centre \\
Trombay, Mumbai, Maharashtra, India, 400085}

\author{A. Pathania}
\affiliation{Astrophysical Sciences Division, Bhabha Atomic Research Centre \\
Trombay, Mumbai, Maharashtra, India, 400085}
\affiliation{Homi Bhabha National Institute, Anushaktinagar, \\
Mumbai, Maharashtra, India, 400094}

\author{S. Kotwal}
\affiliation{Astrophysical Sciences Division, Bhabha Atomic Research Centre \\
Trombay, Mumbai, Maharashtra, India, 400085}

\author{Raj Kumar}
\affiliation{Astrophysical Sciences Division, Bhabha Atomic Research Centre \\
Trombay, Mumbai, Maharashtra, India, 400085}
\affiliation{Homi Bhabha National Institute, Anushaktinagar, \\
Mumbai, Maharashtra, India, 400094}

\author{N. Bhatt}
\affiliation{Astrophysical Sciences Division, Bhabha Atomic Research Centre \\
Trombay, Mumbai, Maharashtra, India, 400085}

\author{K. Chanchalani}
\affiliation{Astrophysical Sciences Division, Bhabha Atomic Research Centre \\
Trombay, Mumbai, Maharashtra, India, 400085}

\author{M. Das}
\affiliation{Astrophysical Sciences Division, Bhabha Atomic Research Centre \\
Trombay, Mumbai, Maharashtra, India, 400085}

\author{K. K. Singh}
\affiliation{Astrophysical Sciences Division, Bhabha Atomic Research Centre \\
Trombay, Mumbai, Maharashtra, India, 400085}
\affiliation{Homi Bhabha National Institute, Anushaktinagar, \\
Mumbai, Maharashtra, India, 400094}

\author{K. K. Gour}
\affiliation{Astrophysical Sciences Division, Bhabha Atomic Research Centre \\
Trombay, Mumbai, Maharashtra, India, 400085}

\author{M. Kothari}
\affiliation{Astrophysical Sciences Division, Bhabha Atomic Research Centre \\
Trombay, Mumbai, Maharashtra, India, 400085}

\author{Nandan Kumar}
\affiliation{Astrophysical Sciences Division, Bhabha Atomic Research Centre \\
Trombay, Mumbai, Maharashtra, India, 400085}

\author{Naveen Kumar}
\affiliation{Astrophysical Sciences Division, Bhabha Atomic Research Centre \\
Trombay, Mumbai, Maharashtra, India, 400085}

\author{P. Marandi}
\affiliation{Astrophysical Sciences Division, Bhabha Atomic Research Centre \\
Trombay, Mumbai, Maharashtra, India, 400085}

\author{C. P. Kushwaha}
\affiliation{Astrophysical Sciences Division, Bhabha Atomic Research Centre \\
Trombay, Mumbai, Maharashtra, India, 400085}

\author{M. K. Koul}
\affiliation{Astrophysical Sciences Division, Bhabha Atomic Research Centre \\
Trombay, Mumbai, Maharashtra, India, 400085}

\author{P. Dorjey}
\affiliation{Department of High Energy Physics, Tata Institute of Fundamental Research, Mumbai, India}

\author{N. Dorji}
\affiliation{Department of High Energy Physics, Tata Institute of Fundamental Research, Mumbai, India}

\author{V. R. Chitnis}
\affiliation{Department of High Energy Physics, Tata Institute of Fundamental Research, Mumbai, India}

\author{R. C. Rannot}
\affiliation{Astrophysical Sciences Division, Bhabha Atomic Research Centre \\
Trombay, Mumbai, Maharashtra, India, 400085}

\author{S. Bhattacharyya}
\affiliation{Astrophysical Sciences Division, Bhabha Atomic Research Centre \\
Trombay, Mumbai, Maharashtra, India, 400085}
\affiliation{Homi Bhabha National Institute, Anushaktinagar, \\
Mumbai, Maharashtra, India, 400094}

\author{N. Chouhan}
\affiliation{Astrophysical Sciences Division, Bhabha Atomic Research Centre \\
Trombay, Mumbai, Maharashtra, India, 400085}

\author{V. K. Dhar}
\affiliation{Astrophysical Sciences Division, Bhabha Atomic Research Centre \\
Trombay, Mumbai, Maharashtra, India, 400085}
\affiliation{Homi Bhabha National Institute, Anushaktinagar, \\
Mumbai, Maharashtra, India, 400094}

\author{M. Sharma}
\affiliation{Astrophysical Sciences Division, Bhabha Atomic Research Centre \\
Trombay, Mumbai, Maharashtra, India, 400085}

\author{K. K. Yadav}
\affiliation{Astrophysical Sciences Division, Bhabha Atomic Research Centre \\
Trombay, Mumbai, Maharashtra, India, 400085}
\affiliation{Homi Bhabha National Institute, Anushaktinagar, \\
Mumbai, Maharashtra, India, 400094}

\begin{abstract}

The radio galaxy NGC\,1275, located at the central region of 
Perseus cluster, is a well known very high energy 
gamma-ray emitter. The Major Atmospheric Cherenkov 
Experiment (MACE) telescope has detected two distinct episodes of 
Very High Energy (VHE, E $>$ 80 GeV) gamma-ray emission from NGC\,1275 
during the period from December 2022 and January 2023. The second outburst, 
observed on January 10, 2023, was more intense of the two, with 
flux reaching 58 \% of the Crab Nebula flux above 80 GeV. The differential 
energy spectrum measured between 80\,GeV and 1.5\.TeV can be described by 
a power-law with a spectral index of $\Gamma = - 2.90 \pm 0.16_{\rm stat}$ 
for both flaring events. The  broadband Spectral Energy Distribution (SED) 
derived from these flares, along with quasi-simultaneous low-energy 
counterparts, suggests that the observed gamma-ray emission can be 
explained using a homogeneous single-zone Synchrotron Self-Compton (SSC) 
model. The physical parameters derived from this model for both flaring 
states are similar. The intermediate state observed between 
two flaring episodes is explained by a lower Doppler factor  or magnetic 
field, which subsequently returned to its previous value during the 
high activity state observed on January 10, 2023.

\end{abstract}

\keywords{Radio Galaxy --- NGC\,1275 --- Very High Energy --- Gamma-Rays  ---IACT --- MACE Telescope}

\section{Introduction} \label{sec:introduction}
NGC\,1275, is a type 1.5 Seyfert galaxy 
\citep{1.5Sey} in the Perseus cluster [z = 0.0176
\citep{Strauss.1992, Falco.1999}]. It resides near the center 
of the large Perseus Cluster of galaxies. This galaxy hosts an active 
galactic nucleus (AGN), and it is classified as a Fanaroff - Riley type I 
radio galaxy based on its radio morphology. This object exhibits an 
extended jet visible in the VLBI images \citep{Nagai.2010}. Unlike blazars, 
the jet of NGC\,1275 is viewed at a larger angle, resulting in relatively 
modest enhancement of the jet core emission due to beaming effects. 
Observing significant gamma-ray luminosities and fast variability in a 
non-blazar AGN, such as NGC\,1275, where a larger viewing angle can cause 
radiation de-boosting, is intriguing and opens up discussions on alternative 
models \citep{Acciari.2009,Aleksic.2014}. Gamma-ray observations and 
studies of the variability of non-blazar AGNs are crucial for 
understanding the location and physical processes responsible for 
extragalactic non-thermal emissions.

In the Very High Energy (VHE) $\gamma$-ray domain, the MAGIC telescope 
first detected NGC\,1275 between 2009 to 2011 in stereoscopic mode
\citep{Aleksic.2010}. Six years later, MAGIC observed NGC\,1275 from
September 2016 to February 2017 and detected a remarkable $\gamma$-ray 
flare on New Years Eve 2016/2017 \citep{Ansoldi.2018}. The peak value of 
the flux was approximately fifty times higher than during the first 
campaign from 2009 to 2011. While the measurements with $Fermi$-LAT 
yielded flux variability on timescales of (1.51 $\pm$ 0.02)\,days 
\citep{BrownAdams.2011}, MAGIC measurements showed marginal flux changes 
on a monthly scale. The VHE band flaring activity from NGC\,1275 
observed during the MACE observation period was also reported by LST-1 
and MAGIC. \citep{ATel.Cortina.2022, ATel.Blanch.2022}.

In this letter, we present the results from observations of NGC\,1275 in 
the VHE band performed with the MACE telescope between December 2022 and 
January 2023, which led to the detection of the source above 80\,GeV on 
two occasions. This paper is structured as follows. The details of 
observations and data analysis procedure are given in Section 
~\ref{sec:observations}. The results obtained with the MACE, 
$Fermi$-LAT and Swift-XRT/UVOT are discussed in 
Section ~\ref{sec:results}. Finally, the discussion and conclusions are 
presented in sections ~\ref{sec:discussion} and ~\ref{sec:conclusions} 
respectively.

\section{Observation Details and Analysis} \label{sec:observations}

\subsection{MACE} \label{subsec:mace}
The observations of NGC\,1275 were carried out using the MACE telescope. 
MACE is a large size (21\,m) ground-based VHE gamma-ray telescope 
\citep{Borwankar.2024} installed at an altitude of 4270\,m above sea 
level at Hanle, in the Ladakh region, a union territory of India. 
NGC\,1275 data used in this paper covers a period of two months, from 
December 2022 to January 2023 in the zenith angle range from -22 degree 
to +30 degree. We focused on 
two datasets corresponding to periods of flaring activity: the night of 
21 December 2022 (referred to as P1) and second on the night of 10 
January 2023 (P2). The intermediate state between these flares is 
designated as P3 hereafter. A total of 5.3 hours of flaring data was 
collected. The analysis cuts applied to NGC\,1275 data were optimized using 
contemporaneous Crab Nebula data and Monte Carlo simulations. The excess 
gamma-ray like events were estimated using the Hillas orientation parameter 
Alpha after gamma-hardon classification and energies were determined 
following the methodology outlined in \citep{Borwankar.2024}.

\subsection{Fermi} \label{subsec:fermi}
The Large Area Telescope (LAT) is a pair conversion $\gamma$-ray detector
on board the $Fermi$-satellite. It surveys the entire sky every 3.2 hours 
and detects $\gamma$-ray photons in the energy range from 20 MeV to $>$ 500\,GeV
\citet{Atwood.2009}. In this work, we have used fully reprocessed
{\it Pass 8} data ($evclass=128$ \& $evtype=3$) from the $Fermi$-LAT
\footnote[1]{https://fermi.gsfc.nasa.gov/ssc/data/access} and the
software package {\it Fermi Science Tools}
\footnote[2]{https://fermi.gsfc.nasa.gov/ssc/data/analysis/software/}
version {\it v10r0p5} with the instrument response functions
$P8R3\_SOURCE\_V3$.

For the source NGC\,1275 (4FGL\,J0319.8+4130), we have extracted the data
within $\rm 10^{\circ}$ region of interest from 17 December 2022 to
16 January 2023 (MJD 59930 -- 59960) in the energy range 0.1 -- 300\,GeV.
To reduce the contamination from the bright Earth limb, we have considered
only the $\gamma$-rays having a zenith angle $\rm < \, 90^{\circ}$. The
unbinned likelihood fitting was performed using the {\it gtlike} task to
extract spectrum and light curve. We have included all the point sources
mentioned in the fourth {\it Fermi}-LAT source catalog (4FGL)
\citep{Acero.2015} within $\rm 10^{\circ}$ radius around the source in
the background model along with the latest galactic diffuse $\gamma$-ray
emission model $gll\_iem\_v07.fits$ and isotropic emission template
$iso\_P8R3\_SOURCE\_V3\_v1.txt$\footnote[3]
{https://fermi.gsfc.nasa.gov/ssc/data/access/lat/BackgroundModels.html}
\citep{Acero.2016}. The position and spectral parameters of NGC\,1275 were
kept the same in the model file as given in the 4FGL catalog and fitted 
with a log-parabola function of the form
$dN/dE = N_{0}(E/E_{b})^{-\alpha\,- \,\beta\, log(E/E_{b})}$ where
$N_{0}$ is the normalization constant, $\alpha$ and $\beta$ 
are the spectral index and curvature parameter, respectively. The spectral 
parameters of NGC\,1275, and the other sources within the $\rm 10^{\circ}$ 
have been set free during likelihood fitting, whereas $E_b$ has been fixed 
to a value 883.6\,MeV throughout the analysis as given in 4FGL. The 
significance of the target source is quantified with the test statistic 
(TS) defined as $TS = -2\,ln(L_{0}/L_{s})$ \citep{Mattox.1996}, where 
$L_{s}$ and $L_{0}$ are the maximum likelihood values for the given model 
with and without target source (NGC\,1275), respectively at that specific 
location.

\subsection{Swift-XRT} \label{subsec:swift-xrt}
X-Ray Telescope (XRT) on board the \textit{Neil Gehrels Swift observatory}
uses grazing incidence Wolter I telescope to focus X-rays onto a CCD detector
with the position accuracy of $\rm 3^{''}$  and operates in the energy
range of 0.3 -- 10\,keV \citep{Burrows.2005}. We have analyzed
quasi-simultaneous $Swift$-XRT data of NGC\,1275, observed during 
01 December, 2022 -- 28 February, 2023. We have used the standard XRTDAS 
package distributed by HEASARC within the HEASOFT package (v.6.24)\footnote[4]
{https://heasarc.gsfc.nasa.gov/lheasoft/download.html}. \textit{XRTPIPELINE}
script has been used to reduce, calibrate, and clean the raw event data files
(Level 1) with the standard filtering criteria\footnote[5]
{https://swift.gsfc.nasa.gov/analysis/xrt\_swguide\_v1\_2.pdf} and using
calibration files of $Swift$ CALDB v.20171113.

NGC\,1275 being an extended source in the X-ray energies
\citep{Fabian.2015, Gallagher.2009}, its spectrum is contaminated
with huge thermal emission from the intracluster medium
\citep{Churazov.2003, Balmaverde.2006}.
For the PC mode data analysis we have chosen the source region within
$\rm 0.4^{'}$ from the center, similar to the procedure described in
\citep{Fukazawa.2018, Ghosal.2020}. Accounting for the extreme
brightness of the source, we have excluded the inner 5 to 6 pixels
(from the core region of source to avoid the pile-up). An annular region
within 25 -- 30 pixels from the center has been chosen as the background 
which includes cluster emission. In WT mode we have chosen the source 
region to be $\rm 0.4^{'}$ from the center and we have used PC mode data 
to estimate the average background in WT mode following the procedure given in
\citep{Fukazawa.2018}. Since the instrument response functions differ in
PC and WT modes in the lower energies, we have estimated WT mode spectra
beyond 1.6\,keV energy.

We extracted the source and background spectra using XSELECT V 2.4e
and generated the ancillary response files (ARFs) using the XRTMKARF command.
We have used the model $\rm ztbabs \star (apec + zpow)$  of XSPECv.12.10.0c 
for the spectral fit and the hydrogen column density was fixed to 
$N_{H}= 1.36 \times 10^{21} cm^{-2}$, which was obtained from HEASARC
website\footnote[6]{https://heasarc.gsfc.nasa.gov/cgi-bin/Tools/w3nh/w3nh.pl}.
The temperature and metal abundance parameters of the apec model were kept 
fixed to the values 4.1 KeV and 0.65, respectively \citep{Fukazawa.2018}. 
The apec normalization values are estimated for each observation ID in PC 
mode (WT mode) and then the apec norm value is frozen to the mean value 
of 0.0232 (0.0489) in PC (WT) mode. The flux and power-law indices for the 
individual observations have been estimated after fixing all the apec 
parameters to the above-mentioned values.

\subsection{Swift-UVOT} \label{subsec:swift-uvot}

The UVOT (Ultraviolet and Optical Telescope) on board the $Swift$
satellite is sensitive in the range of 1600 -- 8000 $\rm A^{\circ}$ and
provides observations in three optical (V,U,B) and three UV (W1,M2,W2)
bands \citep{Roming.2005}. We performed photometry analysis to derive the 
AGN flux. In this measurement, we used image mode data only and combined
multiple frames observed in a single band by applying UVOTIMSUM task.
Using summed images, we selected a circular region of 3 arcsec centred
at the source and another circular region of radius 8 arcsec away
from the source and free from any bright object for background in the
photometric estimation \citep{Imazato.2021}. We then extracted the flux
densities for each band of every observation using the task UVOTSOURCE.
We obtained observed fluxes from the flux densities in each band. Thus, 
the observed fluxes include the contribution of host galaxy components 
and AGN with reddening. Therefore, we need to get the AGN flux after 
minimizing/removing the contribution/effect from other components such 
as the host galaxy and reddening.

The AGN activity is in a compact region and can be similar 
to an isolated star such as a nearby star Pul -3 270315(RA=03:19:41.74244, 
Dec=+41:30:36.688). The radio nature of the NGC\,1275 can be considered 
similar to that of a radio galaxy, such as NGC\,1272. We followed a similar 
method as described in \citep{Imazato.2021} to obtain the AGN contribution.
The AGN contribution was found to be in the range of $\sim$ 70-95\% 
(normalization fraction) in various bands. Thus, obtained AGN fluxes were 
reddening affected. We corrected the interstellar extinction as described 
in \citep{Roming.2009} and \citep{Cardelli.1989} 
using $E_{B-V}=0.1399\pm0.0012$ \citep{Schlafly.2011}. 
The de-reddened value of AGN fluxes are used for SED modeling.

\section{Results}
\label{sec:results}

\subsection{MACE} \label{subsec:mace}
\begin{figure}[ht!]
\centering 
\includegraphics[width=0.48\textwidth]{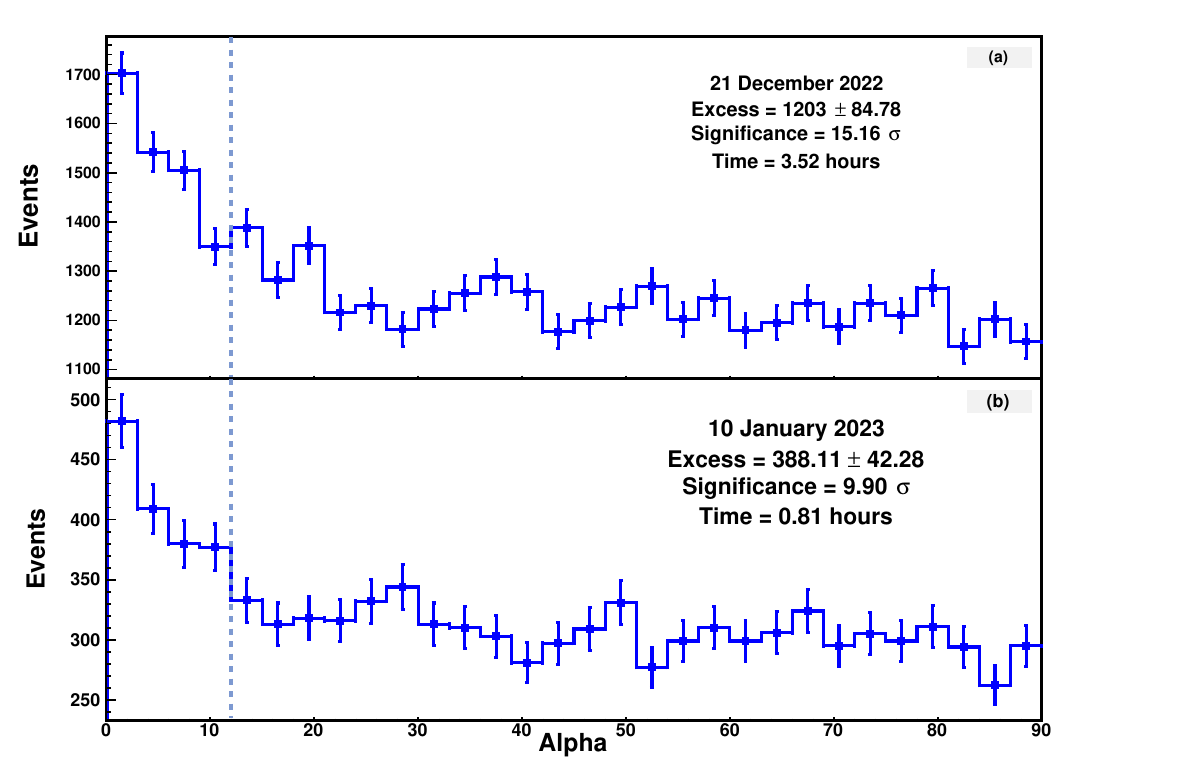}
%\plotone[width=0.48\textwidth]{AlphaCombined.pdf}
\caption{Hillas orientation parameter Alpha distribution of the NGC\,1275 
		(a) 21 December 2022 flare and (b) 10 January 2023 flare detected
		by the MACE telescope. The region between zero and vertical dashed 
        line (at 12.0$^{\circ}$) represents the signal region. 
\label{fig:ngc1275-alpha-plot-combined}}
\end{figure}

For the MACE data between December 2022 and 
January 2023, we performed a source dependent analysis by 
tracking NGC\,1275 at the camera centre. Gamma-ray like events were
selected using the Hillas orientation parameter Alpha 
after passing through 
$\gamma$-hadron classification. Figure \ref{fig:ngc1275-alpha-plot-combined} 
shows the distribution of events 
as a function of the Alpha parameter after application of gamma-domain 
cuts for P1 and P2. The signal was extracted from the Alpha bin of 
12.0$^{\circ}$ while the background region was considered within
27.0$^{\circ}$ $\leq$ Alpha $\leq$ 81.0$^{\circ}$. We detected an
excess of $1203 \pm 84.78$ and 388.11$ \pm 42.28$ $\gamma$-ray like events 
with a statistical significance of 15.16$\sigma$ and 9.9$\sigma$ in 3.52 
and 0.81 hours, respectively. The derived unfolded differential energy 
spectra of the NGC\,1275 for P1 can be described by a simple power-law 
($\chi^{2}$/NDF = 2.96/5) dF/dE = f$_{0} (E/E_{0})^\Gamma$, 
with a photon index $\Gamma = -2.95 \pm 0.1_{ \rm stat}$ and a normalization 
f$_{0}$ = $(5.71 \pm 0.46 ) \times 10^{-11} \rm cm^{-2} \rm s^{-1} \rm TeV^{-1}$ at  
E$_{0} = 400\, \rm GeV$. The differential energy spectra for P2 can also be 
described by power-law model ($\chi^{2}/NDF$ = 1.22/5) with a photon index 
$\Gamma = -2.90 \pm 0.16_{\rm stat}$ and a normalization 
f$_{0}$ = $(8.68 \pm 1.2 ) \times 10^{-11} \rm cm^{-2} \rm s^{-1} \rm TeV^{-1}$ at 
E$_{0} = 400\, \rm GeV$. The integral fluxes derived for NGC\,1275 in the energy 
range of 80 GeV to 1.5 TeV from P1 and P2 are approximately 40\% and 58\% 
of the Crab Nebula flux, respectively. As the MACE observations 
were conducted during the night after those by LST-1 and MAGIC 
\citep{ATel.Cortina.2022, ATel.Blanch.2022}, a direct comparison of flux 
is not possible.
SEDs derived from MACE observations, along with multi-wavelength 
components from $Swift$-UVOT, $Swift$-XRT, and the $Fermi$ gamma-ray 
telescope, were used for emission modeling. 

\subsection{Fermi}
\label{subsec:fermi-results} % used for referring to this section from elsewhere

\begin{figure}[ht!]
\centering 
\includegraphics[width=0.48\textwidth]{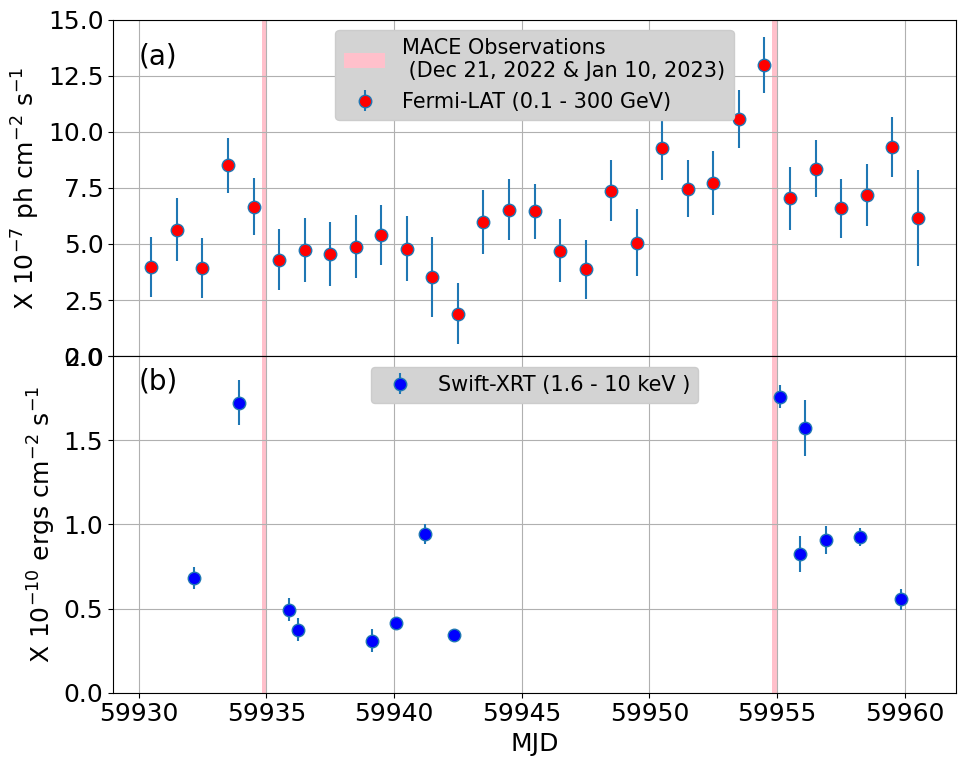}
	\caption{(a) One day binned $Fermi$-LAT light curve of NGC\,1275 in an energy range 0.1 -- 300\,GeV. (b) XRT light curve of NGC\,1275 in the energy range 1.6 - 10.0 KeV
\label{fig:ngc1275-Fermi-Swift-lc}}
\end{figure}

The lightcurve generated using a $Fermi$-LAT data in the 0.1-300\,GeV energy range
is shown in Figure \ref{fig:ngc1275-Fermi-Swift-lc} a. To understand the
spectral evolution we divided the complete interval in seven segments: (1)
MJD 59934 -- 59935, (2) MJD 59936 -- 59940, (3) MJD 59940 -- 59943,
(4) MJD 59943 -- 59948, (5) MJD 59948 -- 59950, (6) MJD 59950 -- 59953
and (7) MJD 59954 -- 59955. The first and the seventh segments correspond
to the contemporaneous observations with MACE. During the first segment, the
photon spectrum in the 0.1-300\,GeV range was described by a log-parabola
distribution, while for the remaining segments, the spectra were well
described by a power-law function. The spectral forms during segments
(2)-(6) primarily differed only in their normalization, with power-law
indices close to 2.

\begin{figure}[ht!]
\centering 
\includegraphics[width=0.48\textwidth]{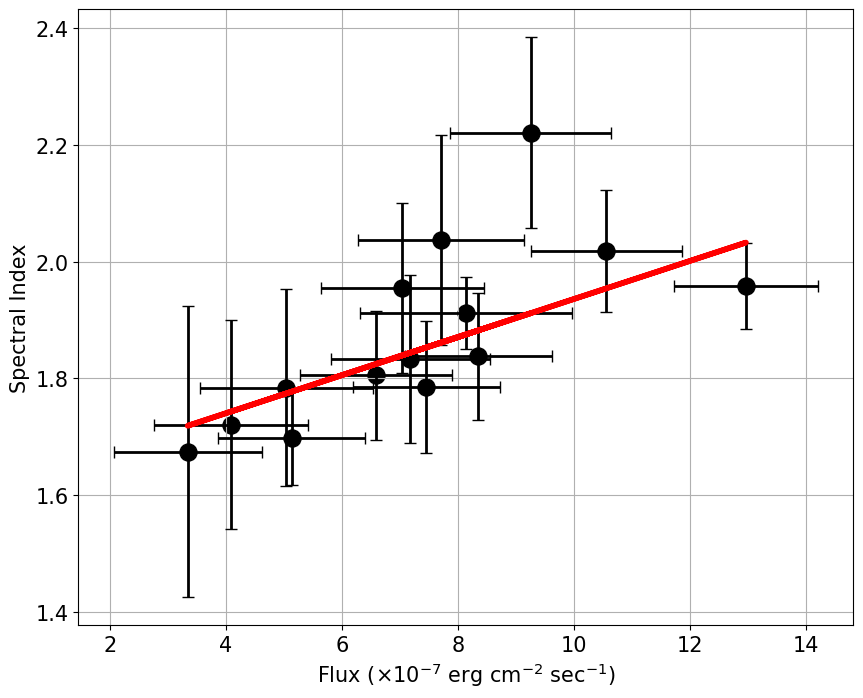}
\caption{Flux-index correlation in the energy range 0.1--300\,GeV during the flaring episodes.
\label{flux-index}}
\end{figure}

The spectral evolution during the flares was analyzed using the correlation 
between flux and spectral index, as illustrated in Figure \ref{flux-index}.
An overall positive correlation was observed, indicating a "softer when 
brighter" behavior. The Pearson correlation coefficient was determined to 
be 0.71, with a null hypothesis probability of 0.4\%. 

\subsection{Swift} \label{subsec:swift-results}

Analysis of $Swift-XRT$ data shows that the source has undergone variations 
in X-ray flux during two flaring episodes detected by the MACE telescope 
(Figure \ref{fig:ngc1275-Fermi-Swift-lc} b). The measured average X-ray flux 
excluding flaring days from the source is 
(6.95 $\pm$ 0.72) $\times 10^{-11} \textbf{\rm ergs\,} \rm cm^{-2} \rm s^{-1}$. During the 
flaring episodes, X-ray flux reached $\sim$ 2.5 times the average flux with 
photon indices being 1.66 $\pm$ 0.11 and 2.07 $\pm$ 0.10, respectively.
Similar index (1.77 $\pm$ 0.17) value was reported during December 31 
2016 - January 01 2017 flare by \citep{Baghmanyan.2017}. The Pearson 
correlation analysis of the flux and indices for the entire duration gives 
correlation coefficient of -0.35 with p value of 0.21, indicating a weak 
anti-correlation between the flux and X-ray photon index.

\section{Discussion}
\label{sec:discussion} % used for referring to this section from elsewhere
The non-thermal emission of blazars in general favours single zone
SSC models 
[see \cite{Mankuzhiyil.2011, Mankuzhiyil.2012} for in-depth study of 
nearby blazar SEDs of Mrk 421 and Mrk 501, and \cite{Tavecchio.2010} for a large
sample SED modeling of Fermi blazars] 
The emission models of all four VHE detected radio galaxies are complex, 
and do not favour a particular class of emission models 
\citep{M87.2012, Centaurus.2018, NGC1275.2018, 3C264.2020}.
Various models (eg: upstream Compton scattering in decelerating jet models, 
spline-sheath layer models, minijets models) have been proposed in the 
case of VHE emission in M\,87  
\citep{Giannios.2010, TavecchioGhisellini.2008, Georganopoulos.2005}. 
In the case of Centaurus\,A, the VHE emission is explained using a sum of 
various Inverse Compton (IC) processes with a dominating share from the 
up-scattering of infrared photons in the dust 
\citep{HESS.Collaboration.2020}. On the other hand SED of 3C\,264
has been modeled using a single zone SSC model with a Doppler
factor of 10 \citep{VERITAS.Collaboration.2020}, which is relatively high 
in such sources. In the case of NGC\,1275, various models have been 
suggested in literature. In order to explain the very first SED that 
include a VHE spectra (together with  other low energy contemporaneous 
data), MAGIC collaboration \citep{Aleksic.2014} employed a single-zone 
SSC model. The parameters used in the model match well with that of other 
blazars, except for a Doppler factor ($\delta$ = 2, 4), as expected in 
the case of radio galaxies. However, in order to explain the fast 
gamma-ray variability detected by MAGIC \citep{Ansoldi.2018} in
2016-2017, various models (like emission from magnetospheric gaps,
relativistic blobs propagating in the jet, external clouds entering the 
jet) were discussed. Fukazawa et al. \citep{Fukazawa.2018} used both 
one-zone and double zone SSC models to explain long term Suzaku and Fermi
observations. In another detailed study using 11 years of data from 
{\it Fermi}-LAT, together with {\it Swift} and AstroSat measurements 
\citep{Gulati.2020}, the emissions during five different flux states of 
the source were explained using a one-zone SSC model. A similar approach 
was used in \cite{Tanada.2018}. 

\begin{figure}[ht!]
\centering 
\includegraphics[width=0.35\textwidth]{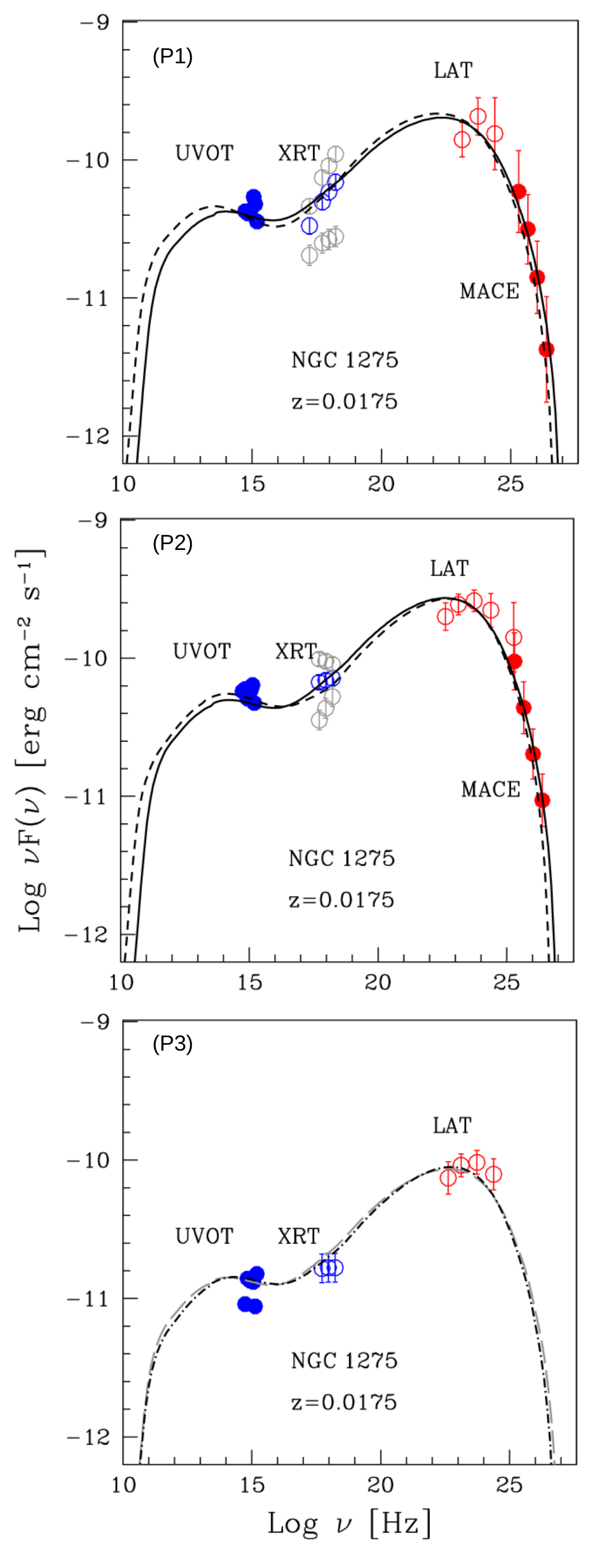}
	\caption{The SED of NGC\,1275 for state P1 is depicted in subfigure (P1). 
	The filled blue symbols represent data from UVOT, while the gray open 
	circles correspond to XRT observations from MJD 59933.9 and 59935.9, with 
	the blue open circles indicating the averaged values. Red open and 
	closed circles represent data from $Fermi$-LAT and MACE, respectively.
	The continuous line represents the one zone SSC model. 
	The SED for P2 is shown in subfigure (P2) using the same symbols as for P1. 
	The SED for the intermediate state P3, measured using UVOT, XRT, and 
	Fermi, is illustrated in subfigure (P3). In P1 and P2 we have also 
	modeled (dashed line) the SEDs by using the parameters suitable for a viewing angle 
	$\theta = 30^{\circ}$. 
\label{fig:ngc1275-sed-P1-P2-P3}}
\end{figure}

The MACE VHE spectra together with other quasi-simultaneous low energy 
flux measurements are fitted with an SSC \citep{Tavecchio.1998} model 
using the method described in \cite{Mankuzhiyil.2011}. Here we assume a 
population of energetic electrons inside a magnetized (B), 
relativistically moving blob (of radius R), emitting through synchrotron 
and SSC mechanisms. The electron energy distribution is described as a
broken power-law, with minimum and maximum Lorentz factors $\gamma_{\rm min}$
and $\gamma_{\rm max}$, respectively and a break at $\gamma_{\rm break}$, with 
indices n$_{1}$ and n$_{2}$, together with normalization factor K. 
The relativistic boosting is encoded in the Doppler factor 
$\delta = [\Gamma(1 - \beta cos\theta)]^{-1}$, where $\Gamma$ is the bulk 
Lorentz factor, $\beta$ is the velocity of the blob in units of the speed 
of light, and $\theta$ is the jet angle relative to the observer's line 
of sight.
In Figure \ref{fig:ngc1275-sed-P1-P2-P3}\,(P1) we have plotted the 
broad-band (eV to TeV) SED of NGC\,1275 during P1, using the 
multi-wavelength data described in Sections \ref{subsec:fermi} and 
\ref{subsec:swift-xrt} and \ref{subsec:swift-uvot}. The VHE spectrum 
of P1 was obtained during MJD 59934.6-59934.8. For the X-ray band, 
we used the average spectrum of MJD 59933.9 and 59935.9, while 
the High Energy (HE) band data was based on one day averaged from 
MJD 59935. 
The optical-UV flux was corrected for the contributions from the 
host galaxy. The obtained SED fit is well in agreement with the 
multiwavelength measurements. Since P2 (like P1) is also a glimpse of 
long term activity of $\sim$ 5 months 
\footnote{https://fermi.gsfc.nasa.gov/ssc/data/access/lat/LightCurveRepository/},
we expect the SSC parameters do not vary drastically. Hence, we attempted to
fit the P2 (VHE [MJD 59954.6] data together with quasi-simultaneous 
Swift [MJD 59955.09 and 59955.89] and LAT [MJD 59955]) with minimal 
changes in the number of SSC parameters with respect to that of P1. 
Interestingly, the observed flux during P2 was reproduced with a minor 
increase ($\sim 30 \%$) in $\gamma_{\rm break}$. This hints that both P1 
and P2 are of a similar state of the radio galaxy in different times.

The parameters of the model during both P1 and P2, are comparable with 
that of other blazar emission models, except for the Doppler factor and 
size of the emission region. However, we note that a lower Doppler factor 
is generally expected in the case of radio galaxies because of the large 
viewing angle. We also note that the recent Multi-wavelength (MWL)
studies of NGC\,1275, provide similar values of $\delta$ to that obtained 
in the present work [$\delta$ =2, 4 \citep{Aleksic.2014}; 3 \citep{Gulati.2020};  
2.7, 3.6 \citep{Tanada.2018}; 2.3, 4.6 \citep{Fukazawa.2018}].

The derived radius in the fit is also roughly an order of magnitude more 
than the radius generally observed in case of blazars. Nevertheless our 
value is comparable with other studies of NGC\,1275 [$\sim$ 1e17 
\citep{Aleksic.2014, Gulati.2020, Tanada.2018}]. Moreover, 
Fermi recorded a HE variability of $\sim$ 5 months during our observation 
period, which can set an upper limit to the radius $R < c t_{var} \delta$.
Using the $\delta=3.5$ from our fit, the upper limit of the radius turns out 
to be $\sim 1\times 10^{18}$ cm. We employ the constraint of 
pair-creation optical depth of TeV photons with soft photons energy, 
to derive a lower limit to the radius \citep{Abdo.2011}. 

\begin{equation} 
\varepsilon _0 = {\delta \, \varepsilon ^{\prime }_0 \over 1+z} \simeq {2 \, \delta ^2 m_e^2 c^4 \over \varepsilon _{\gamma } \, (1+z)^2} \simeq 50 \, \left({\delta \over 10}\right)^2 \left({ \varepsilon _{\gamma } \over {\rm TeV}}\right)^{-1}\,{\rm eV} \ 
\end{equation}                                                                                                                                                                                                                                            

For 1\,TeV photon, the energy of the target soft photon should be $\sim$ 6\,eV. 
The host galaxy subtracted flux at this frequency is 
$\sim 4 \times 10^{-11} \rm erg \hspace{0.2cm } \rm cm^{-2} \rm s^{-1}$. 
The optical depth due to these photons can be written as
\begin{eqnarray} 
\tau _{\gamma \gamma } &\simeq &{\sigma _{\rm T} \, d_L^2 F_0 \, \varepsilon _{\gamma } (1+z) \over 10 \, R \, m_e^2 c^5 \, \delta ^5} \simeq 0.001 \, \left({ \varepsilon _{\gamma } \over {\rm TeV}}\right) \nonumber \\ && \times \left({F_0 \over 10^{-11}\,{\rm erg\;cm^{-2}\;s^{-1}}}\right) \, \left({R \over 10^{17}\,{\rm cm}}\right)^{-1} \left({\delta \over 10}\right)^{-5},\nonumber \\ 
\end{eqnarray}
For $\tau<1$, the radius of the emission 
region $R > 0.75 \times 10^{17}$ cm. It is interesting to note that the 
obtained radius through the SED fit is in agreement with both lower limit 
and upper limit estimations.

To understand the intermediate state P3 between 
P1 and P2 (which was not detected by MACE) we employed Swift (MJD 59939.1) 
and Fermi data (averaged over MJD 59936-59940). We followed a similar 
approach as in the modeling of P2. However, we found that the flux decrease
during the intermediate state is roughly 3 times less than the flux during 
the high activity states in the optical and HE band, where the synchrotron
and IC peak fall respectively. The flux at synchrotron peak is proportional 
\citep{Tavecchio.1998} to
$\nu_{s}F(\nu_{s}) \propto B^{2} R^{3} K \gamma_{b}^{3 - n_{1}} \delta^{4}$
, while that of inverse compton peak is
$\nu_{c}F(\nu_{c}) \propto B^{2} R^{4} K^{2} \gamma_{b}^{2(3 - n_{1})} \delta^{4}$.
Hence, we can rule out change in K, R and $\gamma_{\rm break}$
as a major reason for the decrease in the flux (if the other parameters 
are not varying). So, we fixed all parameters as constant, and left 
$\delta$ as varying, to approximate the intermediate observation with the 
model. We found that a $\delta = 2.45$ can broadly reproduce the low 
flux state. At the same time, a slight change in 
$\gamma_{\rm break} = 1.9 \times 10^4$ will further fine tune the 
approximation.
Row 3 of Table~\ref{table:parameters} presents the parameters for 
the intermediate state (P3) with varying Doppler factor. Additionally, 
we repeated the analysis to achieve a similar fit by adjusting the 
magnetic field to B = 10.0 mG and $\gamma_{\rm break} = 1.7 \times 10^4$, 
as detailed in row 4 of Table~\ref{table:parameters}.
Hence, we can summarize the activity 
in December 2022 - January 2023 as follows. A change in magnetic field 
(20 to 10 mG) or Doppler factor (3.5 to 2.45), with a minor change in 
$\gamma_{\rm break}$ decreased the flux from P1 state to intermediate state. 
The same parameter increased to the previous value (with small difference 
in $\gamma_{\rm break}$) shifted the source to the P2 state.

\begin{figure}[ht!]
\centering	
\includegraphics[width=0.60\textwidth]{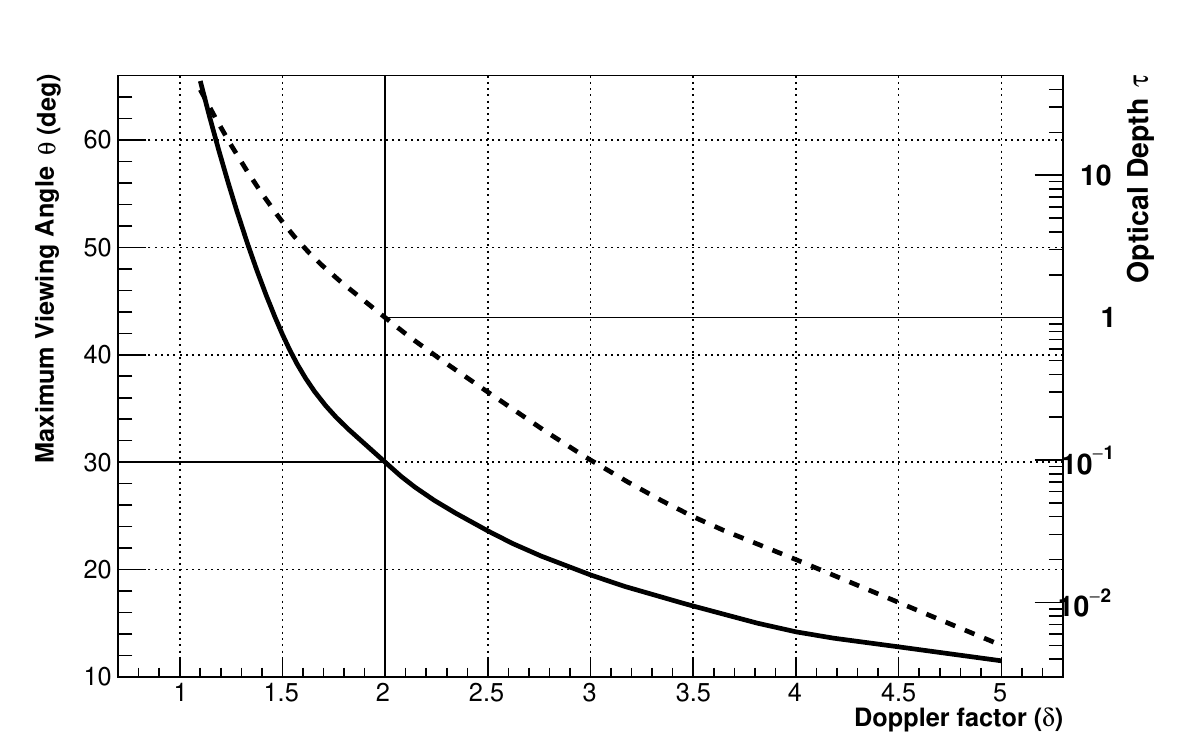}
%\plotone[width=0.48\textwidth]{AlphaCombined.pdf}
	\caption{The maximum viewing angle (solid line) and optical depth (dotted line) for different Doppler factor values, considering a $\sim 5$ months variability period. 
	The black thin lines mark $\tau$ = 1 at $\theta$ = 30$^{\circ}$.
\label{fig:ngc1275-vangle}}
\end{figure}

As stated before, the obtained $\delta=3.5$ during MACE detection 
was notably smaller than that of blazars, which corresponds to a maximum 
viewing angle $\theta \sim 17^{\circ}$ with a lorentz factor $\Gamma=3.5$. 
This viewing angle is consistent with the previous studies 
[eg: $\theta = 15^{\circ}$ for $\delta=4$ \citep{Aleksic.2014}; $20^{\circ}$ 
\citep{Gulati.2020}]. To estimate the maximum possible viewing angle during 
MACE observations, we sought the minimum acceptable $\delta$ and 
corresponding size of the emission region (R) considering the $\sim 5$ 
months variability. For each $\delta$-R combination, we calculated the 
optical depth within the source, ensuring $\tau<1.$. 
Figure \ref{fig:ngc1275-vangle} illustrates the variation of maximum viewing 
angle and optical depth as a function of Doppler factor. 
The minimum allowed $\delta$ was found to be 2. This 
corresponds to a maximum viewing angle $\theta=30^{\circ}$. Lower values 
of $\delta$ (i.e $\theta > 30^{\circ}$) and a maximum allowed R would result 
in higher $\tau$ making 
the medium optically thick for VHE $\gamma$-rays. We have also modeled 
[shown in dashed lines in Figure \ref{fig:ngc1275-sed-P1-P2-P3} 
(P1 and P2), and Table \ref{table:parameters} (row 5 and 6)] the SEDs during MACE 
detections, using these parameters (ie, $\delta=2$ and R=$1\times 10^{18}$ cm). 
In Figure \ref{fig:ngc1275-vangle} we show $\theta$ (left vertical axis), and $\tau$ 
(right vertical axis), corresponds to different $\delta$ values. While the obtained 
maximum viewing angle is in agreement with the value (11$^{\circ}$) obtained 
by \cite{Lister.2009} through radio observations and marginally overlaps 
with the viewing angle 
estimates of \cite{Walker.1994} ($\theta=30^{\circ}-55^{\circ}$), it
disagrees with the values ($\theta=60^{\circ} \pm 16^{\circ}$) obtained 
by \cite{FujitsNagai.2017}.
  
  %This discrepancy between high-energy observations (in the framework of one-zone emission model) and radio observations could be attributed to jet precision issues in radio galaxy jets.

Spine-sheath layer model is often used to explain the emission from radio 
galaxies \citep{TavecchioGhisellini.2014, Rieger.2018} at higher viewing 
angles, where different co-spatial layers of the jet advance at different 
bulk velocities. However, this model requires very small $\Gamma$ at the 
layer where high energy emission is produced.  This results in very small 
$\delta$ at large $\theta$ leading to a high optical depth ($\tau >> 1.$). 
Therefore this model would not be suitable for scenarios  where VHE 
spectra extend beyond 1\,TeV. On the other hand, magnetospheric models 
\citep{NeronovAharonian.2007, Katsoulakos.2018}, predict particle 
acceleration at high magnetic fields near the blackhole horizon scales. 
Considering the blackhole mass \citep{Scharwachter.2013} of NGC\,1275, 
magnetospheric models can not account for flaring activities beyond 
day-time scales.

%\begin{deluxetable*}{cccclDlcccc}
\begin{deluxetable*}{ccccccccccc}[ht!]
\tablenum{1}
\tablecaption{Single-zone SSC Model Parameters for the NGC\,1275 data-set \label{table:parameters}}
\tablewidth{0pt}
\tablehead{
\colhead{Epoch} & \colhead{$\gamma_{min}$} & \colhead{$\gamma_{break}$} & \colhead{$\gamma_{max}$} & \colhead{n1} & \colhead{n2} & \colhead{K} & \colhead{B} & \colhead{R} & \colhead{$\delta$} \\
\colhead{} & \colhead{} & \colhead{$\times 10^{4}$} & \colhead{$\times 10^{6}$} & \colhead{} & \colhead{} & \colhead{$\times 10^{5} \rm cm^{-3}$} & \colhead{mG} & \colhead{$\times 10^{17} \rm cm$} & \colhead{}
}
%\decimalcolnumbers
\startdata
P1 & 30 & 1.13 & 2.60 & 2.36 & 3.45 & 1.05 & 20.0 & 1.90 & 3.5  \\
P2 & 30 & 1.45 & 2.60 & 2.36 & 3.45 & 1.05 & 20.0 & 1.90 & 3.5  \\
P3 (change in $\delta$) & 30 & 1.90 & 2.60 & 2.36 & 3.45 & 1.05 & 20.0 & 1.90 & 2.45  \\
P3 (change in B) & 30 & 1.70 & 2.60 & 2.36 & 3.45 & 1.05 & 10.0 & 1.90 & 3.5  \\
\hline
P1($\theta$ = 30$^{\circ}$) & 120 & 1.30 & 2.60 & 2.36 & 3.45 & 0.18 & 13.0  & 10 & 2  \\
P2($\theta$ = 30$^{\circ}$) & 25 & 2.3 & 2.60 & 2.36 & 3.45 & 0.14 & 13.0  & 10 & 2  \\
\enddata
\end{deluxetable*}

\section{Conclusions}
\label{sec:conclusions} % used for referring to this section from elsewhere
In this work we present the detection of the episodic activity of 
NGC\,1275 measured from December 2022 to January 2023 with MACE 
telescope. We found two nights during which NGC\,1275 was in high flux 
state compared to quiescent state.  We also analysed quasi-simultaneous 
low energy data to understand the emission processes during the days of 
MACE detection. Employing single zone SSC model, we analyzed the SEDs 
observed across distinct epochs labeled  P1, P2 (flaring state), and an 
intermediate phase, P3. This approach helped us to understand the 
variability characterizing emission from NGC\,1275. It was revealed by 
our analysis that the physical parameters during the flaring state P1 and 
P2 were almost identical except for a small difference in the break 
energy of the electron energy distribution ($\gamma_{\rm{break}}$). The 
derived size of the emission region is in agreement with the previous 
studies and the limits obtained from the gamma-ray variability and optical 
depth. We have also found that the intermediate state P3 was a result 
of lower Doppler factor or magnetic field (with a small 
difference in $\gamma_{\rm{break}}$). 
We have also estimated the maximum possible viewing angle of the 
jet (considering an optical depth $\tau <$ 1 for VHE spectrum), 
which is found to be 30$^{\circ}$. In conclusion, the detailed analysis 
of the VHE gamma-ray emission has provided valuable insights into emission 
mechanisms in the radio galaxy NGC\,1275. The detection of flaring 
activity not only highlights the dynamic nature of NGC\,1275 but also 
emphasizes the importance for continuous and comprehensive monitoring with 
ground based gamma-ray telescopes. Such observations are crucial for 
advancing our understanding of both NGC\,1275 in particular and radio 
galaxies in general.

\begin{acknowledgments}
We express our sincere gratitude to Dr. A. K. Mohanty, Chairman Atomic
Energy Commission and Secretary Department of Atomic Energy, Shri Vivek
Bhasin, Director, Bhabha Atomic Research Centre, and Dr. S. M. Yusuf,
Director Physics Group, Bhabha Atomic Research Centre for their guidance,
continuous support and encouragement.
\end{acknowledgments}

%\appendix


\bibliography{sample631}{}
\bibliographystyle{aasjournal}



\end{document}